\documentclass[article,moreauthors,pdftex]{Definitions/mdpi} 
\firstpage{1} 
\pubvolume{1}
\issuenum{1}
\articlenumber{0}
\pubyear{2021}
\copyrightyear{2021}
\externaleditor{Academic Editor: Enrique Maciá
Barber} 
\datereceived{10 May 2021} 
\dateaccepted{16 June 2021} 
\datepublished{} 
\hreflink{https://doi.org/} 
\pdfoutput=1

\Title{Assessing bound states in a one-dimensional topological superconductor: Majorana versus Tamm}

\TitleCitation{Assessing bound states in a one-dimensional topological superconductor: Majorana versus Tamm}

\Author{Lucia Vigliotti $^{1}$, Fabio Cavaliere 
$^{1,2}$, Matteo Carrega $^{2}$ and Niccolò Traverso 
Ziani $^{1,2}$*}

\AuthorNames{Lucia Vigliotti, Fabio Cavaliere, Matteo Carrega and Niccolò Traverso Ziani}

\AuthorCitation{Vigliotti, L.; Cavaliere, F.; Carrega, M.; Traverso Ziani, N.}

\address{%
$^{1}$ \quad Dipartimento di Fisica Università di Genova, Via Dodecaneso 33, 16146 Genova (Italy)\\
$^{2}$ \quad CNR-Spin, Via Dodecaneso 33, 16146 Genova (Italy)}

\corres{Correspondence: traversoziani@fisica.unige.it}

\abstract{Majorana bound states in topological superconductors have attracted intense research activity in view of applications in topological quantum computation. However, they are not the only example of topological bound states that can occur in such systems. We here study a model in which both Majorana and Tamm bound states compete. We show both numerically and analytically that, surprisingly, the Tamm state remains partially localized even when the spectrum becomes gapless. Despite this fact, we demonstrate that the Majorana polarization shows a clear transition between the two regimes.}

\begin{document}
\nolinenumbers
\section{Introduction}
Non abelian quasiparticles have attracted intense theoretical and experimental research activities in the last years, in view of their potential applications in topologically protected quantum computation \cite{das1}. Several proposals for the experimental realization or detection of such exotic states of matter in interacting systems have hence been put forward \cite{para1,teo1,para2,para3,para4,para5,teo2,teo3,df1,addd1,addd2,addd3}. Despite steady progress, the need to enforce the coexistence of superconductivity and electronic interactions makes their experimental realization challenging \cite{suphall1,suphall2,suphall3}. In the absence of interactions, the possibility to engineer non abelian Majorana zero modes \cite{majo1} in spinless p-wave superconductors has been 
predicted \cite{majo2,majo3}. The unambiguous experimental realization of such states is again far from straightforward, due to the fact that spinless p-wave superconductivity is very rare in nature. However, it has been realized that this kind of pairing could be achieved by means of proximity induced superconductivity on semiconducting structures with pronounced spin dependent properties. One of the first example of a system that, once proximitized, can host Majorana bound states, is represented by topological insulators \cite{fu1,fu2,cb1}, where strong spin-orbit coupling gives rise to inverted bands behaviour. However, it additionally requires the presence of magnetic barriers to allow for Majorana bound state formation. The difficulty in implanting magnetic barriers has however slowed down the development of such a platform. Recently, the possibility to employ quantum point contacts \cite{qpc} instead of magnetic barriers has been considered, providing a promising alternative for topological insulator-based Majorana platforms \cite{qpc1,qpc2}. Several other platforms have been proposed, ranging from ferromagnetic atomic chains with superconducting pairing \cite{fc}, planar Josephson junctions in 2D semiconductors \cite{pjj,teo5,teo6}, and the more popular setup based on spin-orbit coupled quantum wire with proximity superconductivity and applied magnetic fields \cite{mf11,mf12}.

In the last case, several experiments have reported signatures compatible 
with the presence of Majorana zero modes \cite{aguado}. However, a clear evidence of the quantization of the zero bias peak, which is one of the predicted signatures ascribed to the presence of Majorana zero modes, or other unambiguous fingerprints of their formation are still lacking to date. Indeed, an additional challenge in the field is posed by the fact that Majorana zero modes behave, in many terms, similarly to more conventional Andreev bound 
states, which do not exhibit non-abelian statistics, and are thus less useful in quantum computation
\cite{abs0,abs1,abs2,abs3,abs4,abs5,abs6,abs7,abs8,abs9,abs10,abs11,abs12,abs13,abs14,abs15,abs16,abs17}. Indeed, on one hand it is still possible to distinguish between trivial Andreev bound states that emerge at finite chemical potential in clean systems and Majorana bound states due to the fact that the phase containing Majorana fermions and the one with the trivial bound states are separated by a region with extended states only \cite{gbu1}. On the other hand, the distinction becomes more subtle in the presence of disorder, spatial variations of the confinement potentials, or whenever the formation of quantum dots \cite{gbu2} takes place. A detailed analysis of the scenarios in which Majorana fermions and other bound states coexist is hence in order.\\

\noindent In this work, we concentrate on this aspect by studying a simple model where different bound states may coexist. To this end, we consider a one-dimensional finite size spinless p-wave superconductor in the presence of a spatially-periodic modulation of the local chemical potential. The system is numerically investigated by means of extensive exact diagonalization. We find that the bulk of the system is always gapped at the chemical potential, except for a gapless point that marks the boundary between two phases dominated either by the periodic local potential (A) or the superconducting one (B). In the A phase either a bound state is localized at one end of the system or no boundary states are present, depending on the phase of the modulated potential. The possible boundary state is adiabatically connected with the system in the absence of superconductivity and is hence qualitatively speaking a Tamm state \cite{tamm1,tamm2,tamm3}. In the B phase, on the other hand, one recovers the more conventional Majorana scenario, with the zero mode split into two Majorana modes at both ends of the system. Surprisingly, at the transition between A and B the Tamm state remains partially localized. This behavior poses the question if the Majorana fermions in the B phase are still topological in nature or if the reminiscence of the Tamm state eventually spoils their properties.\\
\noindent In the second part of the work we answer this question, deriving and analyzing a linearized low-energy model which can be analytically solved. This allows us to interpret and discuss the physics observed in the first part in terms of Goldstone-Wilczek charges \cite{gw1,gw11,gw2,gw3,gw4,gw5,gw6,tamm2,gw8}.\\
Finally, to elucidate the interplay between the different kinds of bound states, and to possibly discriminate between them, we evaluate the Majorana polarization \cite{mp1,mp2,mp3} associated to the boundary states. This quantity is one of the standard tools introduced to characterize topological superconductors. This allows us to conclude that the phase B is indeed characterized by the presence of Majorana fermions.\\

\noindent In more details, the article is divided as follows. In Sec. 2 we 
analyze the results for the p-wave superconductor model, in Sec. 3 we derive the exactly solvable model and discuss it. Sec. 4 contains the calculation of the Majorana polarization. Our conclusions are finally drawn in Sec. 5.

\section{The quadratic model}
\subsection{Hamiltonian}
The system we inspect is a finite-size, spinless one-dimensional p-wave superconductor in the presence of an additional periodic potential \cite{bernevig}. More specifically the Hamiltonian, defined on a segment of length $L$, is
\begin{equation}
H=\frac{1}{2}\int_0^L \Psi^\dag(x)\mathcal{H}(x)\Psi(x)dx,   
\end{equation}
with the Bogoliubov-de Gennes (BdG) Hamiltonian density $\mathcal{H}(x)$ given by ($\hbar=1$)
\begin{equation}
 \mathcal{H}(x)= \left(\frac{-\partial_x^2}{2m}-\mu_0-\mu_1(x)\right)\tau_3+(-i)\frac{\Delta}{k_F}\partial_x\tau_1,
\end{equation}
and $\Psi(x)=(\psi(x),\psi^\dag(x))^T$, with $\psi(x)$ a fermionic annihilation field. We impose open boundary conditions, implying $\Psi(0)=\Psi(L)=0$. In the Hamiltonian, $m$ is the effective mass, $\mu_0$ sets the filling with\footnote{Note that, in view of the finite size of the system, we also impose $k_F=n_F\pi/L$, with $n_F$ a positive integer.} $k_F=\sqrt{2m\mu_0}$, $\Delta$ the strength of the p-wave superconducting pairing and $\tau_i$ ($i=1,2,3$) are the Pauli matrices. For future use we also introduce $\tau_0$ as the 2X2 identity matrix. Finally we set
\begin{equation}
    \mu_1(x)=-B\cos(2k_F x+\varphi),
\end{equation}
with the phase $\varphi$ kept as a free parameter. Here, $\mu_1(x)$ is a periodic potential, that can emerge due to external gates \cite{joha} or via a coupling to the phonons \cite{peierls}. The parameter $B$ parametrizes the strength of the term, while the phase $\varphi$, which is clearly an irrelevant parameter when periodic boundary conditions are imposed, becomes essential in the case of open boundary conditions. The manipulation of the phase $\varphi$ can be envisioned if $\mu_1(x)$ is due to external finger gates \cite{joha}. In this case, the spatial variations of the potential can be fully manipulated.\\
The energy spectrum and the wavefunctions are evaluated via a numerical exact 
diagonalization procedure, which consists of expanding the eigenfunctions 
$\psi_{\nu}(x)$ of the Hamiltonian on the basis of the states of a free particle in a infinite 1D box of length $L$
\begin{equation}
\psi_\nu(x)=\sqrt{\frac{2}{L}}\sum_n \sin\left(\frac{n\pi x}{L}\right) c_n(\nu)
\end{equation}
and diagonalizing the associated matrix of the Hamiltonian on this basis. 
This allows us to obtain the energy spectrum $E_\nu$ and the weights $c_n(\nu)$ of the eigenfunctions of the problem. It is here worth to mention that since we adopt the Bogoliubov-de Gennes formalism we artificially double the spectrum by introducing a redundant chiral symmetry, and hence only half of the eigenfunctions of the Hamiltonian matrix have physical meaning. Indeed, by only taking the positive energy eigenstates one recovers the correct excitation spectrum. The diagonalization is performed in Mathematica (TM) and up to 450 box states per BdG sector are employed depending on the values of $B$ and $\Delta$, to ensure numerical convergence with a relative error $\delta\lesssim 10^{-3}$ on the energy spectrum.
\subsection{Tamm states}
We begin our analysis by considering the case $\Delta=0$. 
\begin{figure}[htbp]	
\widefigure
\includegraphics[width=14 cm]{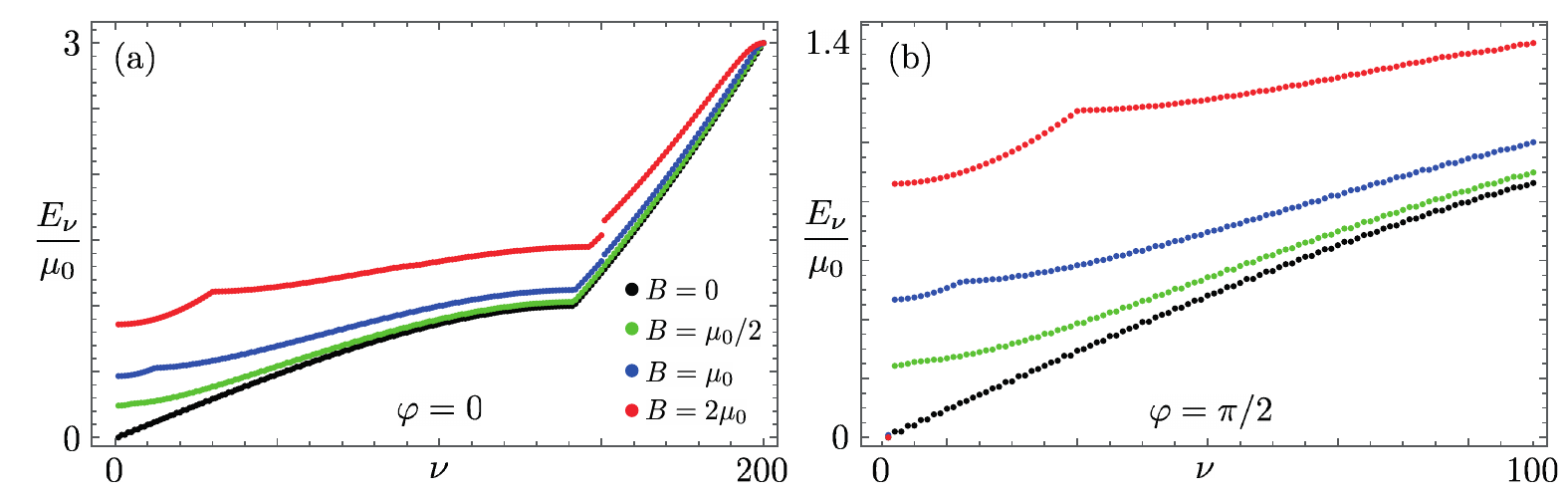}
\caption{Plots of the energy spectrum for different values of the amplitude of the periodic potential: $B=0$ (black); $B=\mu_0/2$ (green); $B=\mu_0$ (blue); $B=2\mu_0$ (red). The case $\varphi=0$ is shown in panel (a) for the first 200 eigenstates, while $\varphi=\pi/2$ in panel (b) for the first 100 eigenstates.\label{fig1}}
\end{figure}  
Figure \ref{fig1} shows the energy spectrum for different values of the strength $B$ and different phases $\varphi$. As a general feature we observe that for $B\neq 0$ the periodic potential opens a gap at zero energy -- where the  chemical potential is set. Also, secondary gaps can occur when the amplitude $B$ becomes large, $B\gtrsim\mu_0$. This additional gap, visible in \ref{fig1}(a) for $\nu\sim 150$, opens, having in mind periodic boundary conditions and a folded scheme for taking into account the periodic perturbation, at $k=0$. For the same magnitudes of $B$, an additional feature appears for small $\nu$ (see the blue and red curves). This is due to the change in nature of the lowest energy states, which, for $B>\mu_0$, are dominated by the periodic potential. Comparing Figs.\ref{fig1}(a) 
and (b) one can observe that for {\em almost} all the states, the overall 
shape of the energy spectrum depends weakly on the phase $\varphi$, being 
almost insensitive to it for $E\gtrsim\mu_0$. However, inspecting the case $\varphi=\pi/2$ one can observe that a zero-energy mode occurs within 
the gap, which suggests that the low-energy sector of the spectrum may exhibit a more pronounced dependence on $\varphi$.
\begin{figure}[htbp]	
\widefigure
\includegraphics[width=14 cm]{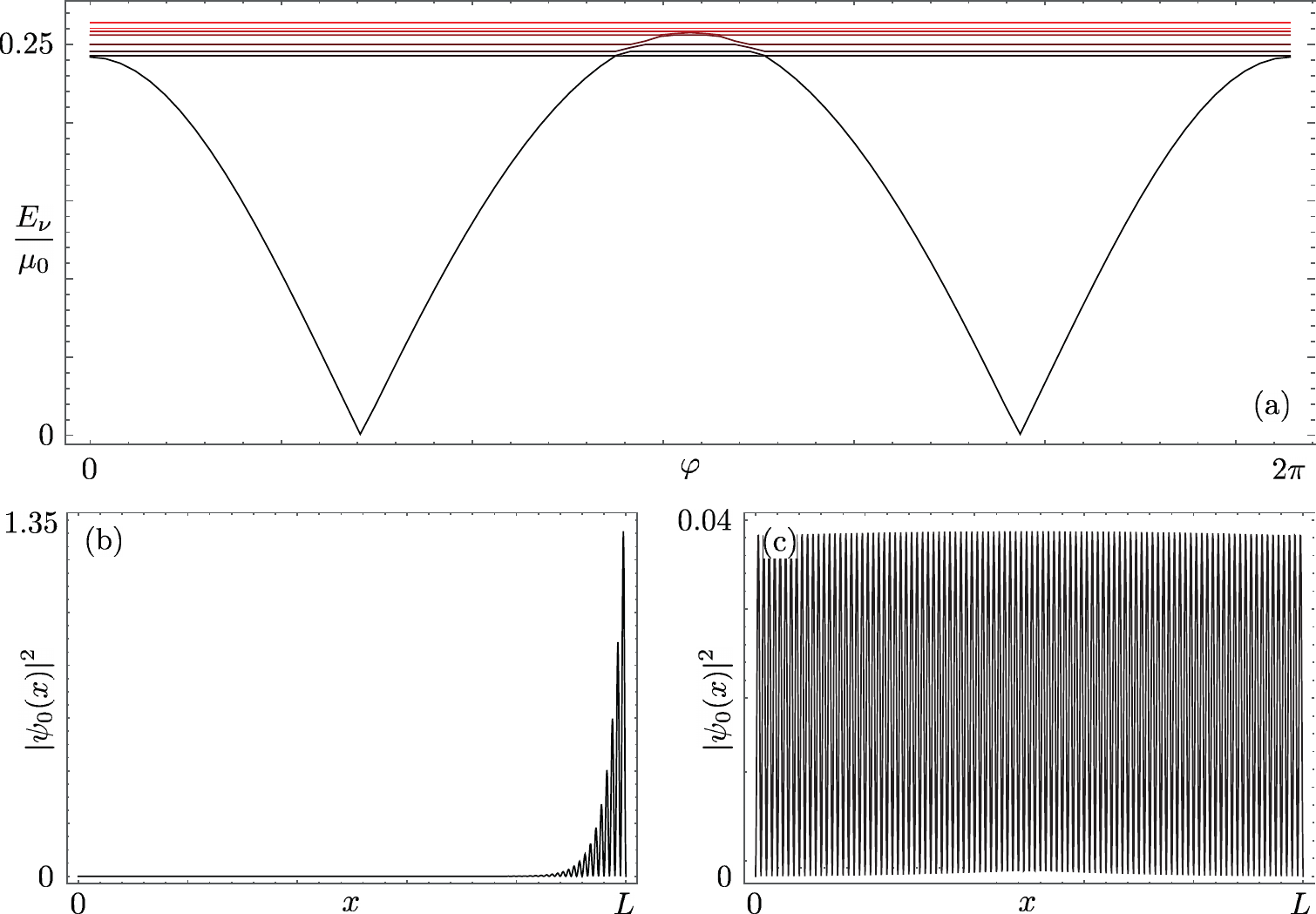}
\caption{Plot of $E_{\nu}/\mu_0$ for $0\leq\nu\leq 7$ (from black to red) 
as a function of $\varphi$ (panel a) and probability density of the $\nu=0$ state $|\psi_0(x)|^2$ (units $L^{-1}$) as a function of $x$ for $\varphi=\pi/2$ (panel b) and $\varphi=0$ (panel c). In all panels, the amplitude of the periodic potential is set at $B=\mu_0/2$.\label{fig2}}
\end{figure}
To check this fact, Fig. \ref{fig2}(a) shows the behaviour of the first eight eigenstates with $E_\nu\geq 0$ as a function of $\varphi$ for the representative case of $B=\mu_0/2$. As one can clearly see, the lowest energy state exhibits a quite dramatic dependence on the phase, becoming a zero mode for $\varphi=(2p+1)\pi/2$ with $p$ an integer. The wavefunction of such a zero-energy state is shown, for $\varphi=\pi/2$ in Fig. \ref{fig2}(b) and it corresponds to a localized state, which in this context is usually referred to as a Tamm state. Such state is 
respectively localized either at the right ($\varphi=\pi/2+2\pi p$) or at the left ($\varphi=3\pi/2+2\pi p$) edge of the system. Such Tamm state is starkly different from the delocalized states obtained at $\varphi=\pi p$, which extend over the entire length of the system -- see the representative example reported in Fig. \ref{fig2}(c) for the case $\varphi=0$. In general, for $\varphi\neq \pi p$ a bound state within the gap always occur, with possibly a non-zero energy and a larger localization length.
\subsection{Majorana states}
The regime $B=0$ with $\Delta\neq 0$ is more known since it represents the paradigmatic model that hosts Majorana bound states: even in this case a gap opens around the chemical potential and a fermionic zero-energy mode appears in the spectrum. 
\begin{figure}[htbp]	
\widefigure
\includegraphics[width=14 cm]{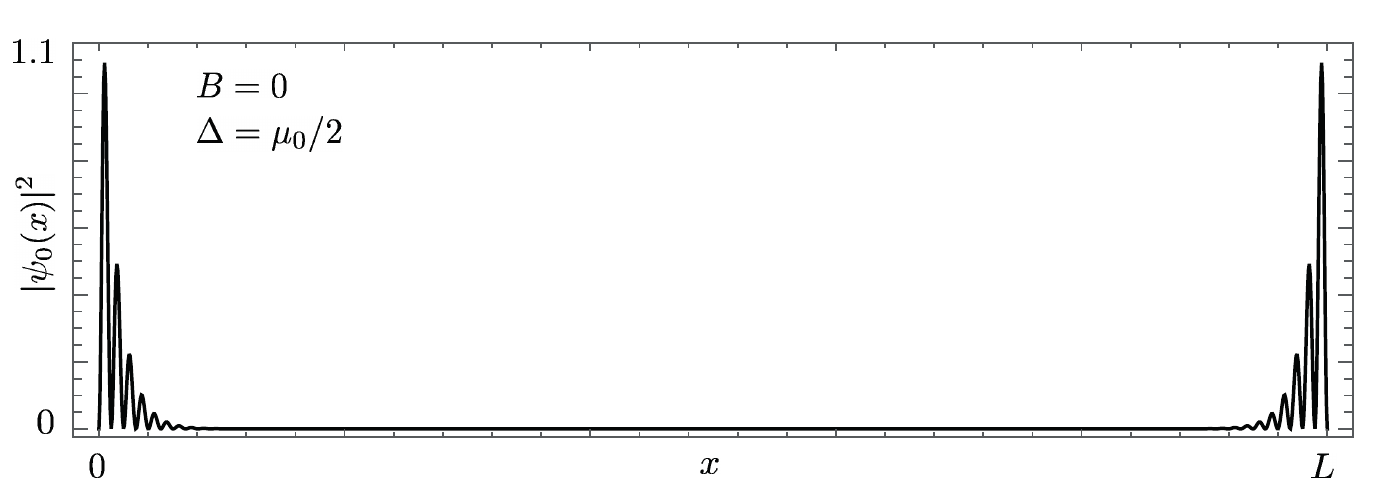}
\caption{Probability density of the zero-energy Majorana state $|\psi_0(x)|^2$ (units $L^{-1}$) as a function of $x$ for $B=0$ and $\Delta=\mu_0/2$.\label{fig25}}
\end{figure}  
The BdG wave function corresponding to this fermionic state is significantly non-zero close to both ends of the segment -- see Fig. \ref{fig25}. Formally, one can interpret this non-local fermionic state as two local Majorana states, each one localized close to one end only. This is obviously meaningful only when the localization length is smaller than the length of the system.
\subsection{Competition between Tamm and Majorana states}
Let us now address the case when both $B$ and $\Delta$ are non-zero and thus the two gaps induced by the oscillating potential and by the p-wave pairing compete. We are particularly interested into the interplay between the two (qualitatively different) localized Tamm and Majorana states and thus we set from now on $\varphi=\pi/2$ to achieve the most dramatic effects. Indeed, $\varphi=\pi/2$ corresponds to a maximally localized Tamm state. As a general feature, we find that in this case, for any value of $B,\,\Delta$ a zero-energy mode occurs.
\begin{figure}[H]	
\widefigure
\includegraphics[width=14 cm]{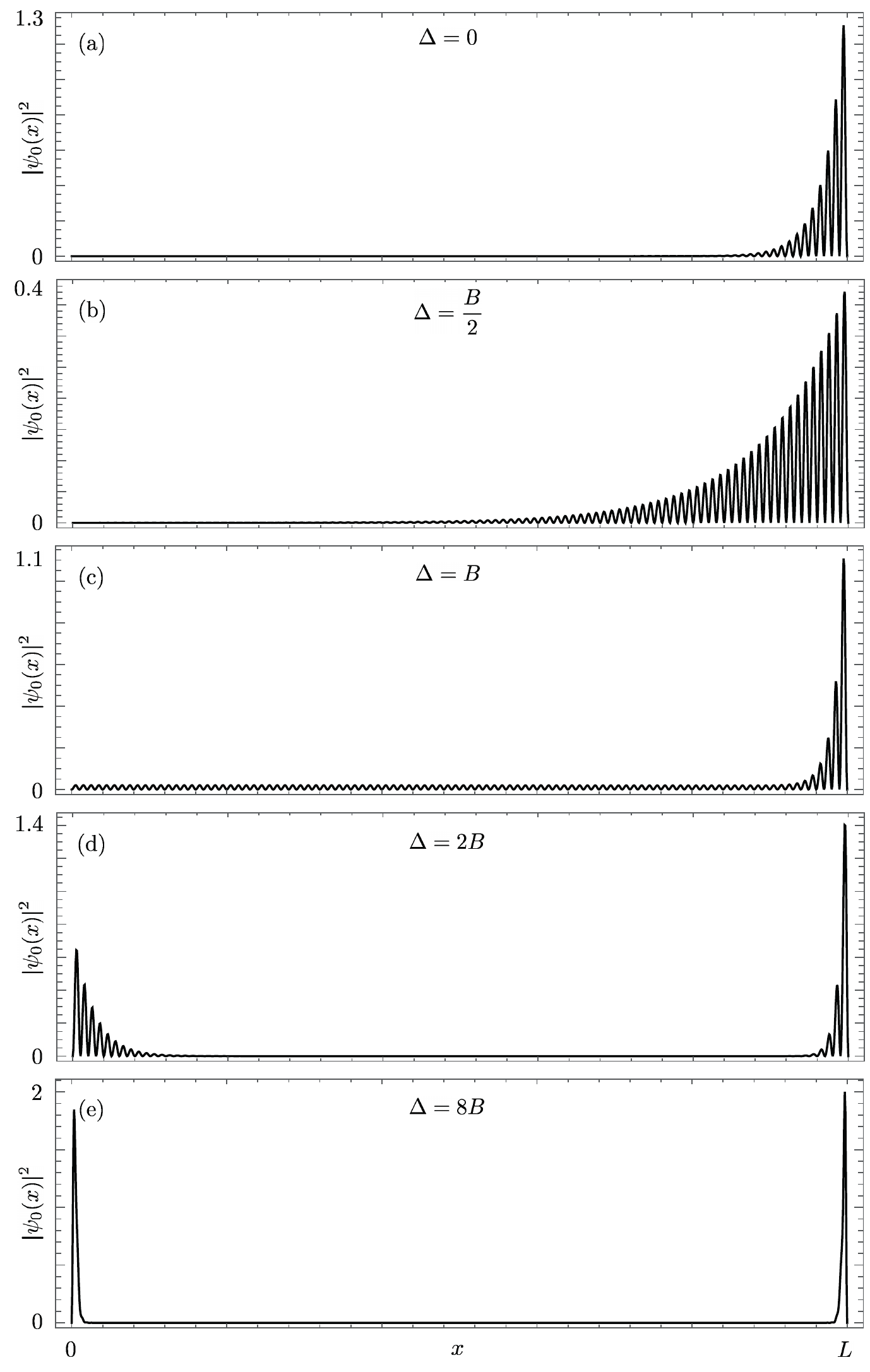}
\caption{Probability density of the zero-energy state $|\psi_0(x)|^2$ (units $L^{-1}$) as a function of $x$ for different values of the p-wave pairing strength: $\Delta=0$ (a); $\Delta=B/2$ (b); $\Delta=B$ (c); $\Delta=2B$ (d); $\Delta=8B$ (e). In all panels, $B=\mu_0/4$ and $\varphi=\pi/2$.\label{fig3}}
\end{figure}  
Figure \ref{fig3}(a-e) shows the probability amplitude $|\psi_0(x)|^2$ of this zero-energy mode for $B=\mu_0/4$ and increasing values of the superconducting pairing. As $\Delta$ is increased from zero, the localized state broadens -- see panel (b). However, when $\Delta=B$ -- see panel (c) -- a peculiar phenomenon occurs: the former Tamm state gets once again sharper while the wavefunction in the rest of the system becomes non-zero 
and with a flat envelope. When $\Delta>B$, see panel (d), a second localized state develops. Finally, when $\Delta\gg B$ the wavefunction tends to 
recover a symmetric shape localized at both ends of the system.
\begin{figure}[H]	
\widefigure
\includegraphics[width=14 cm]{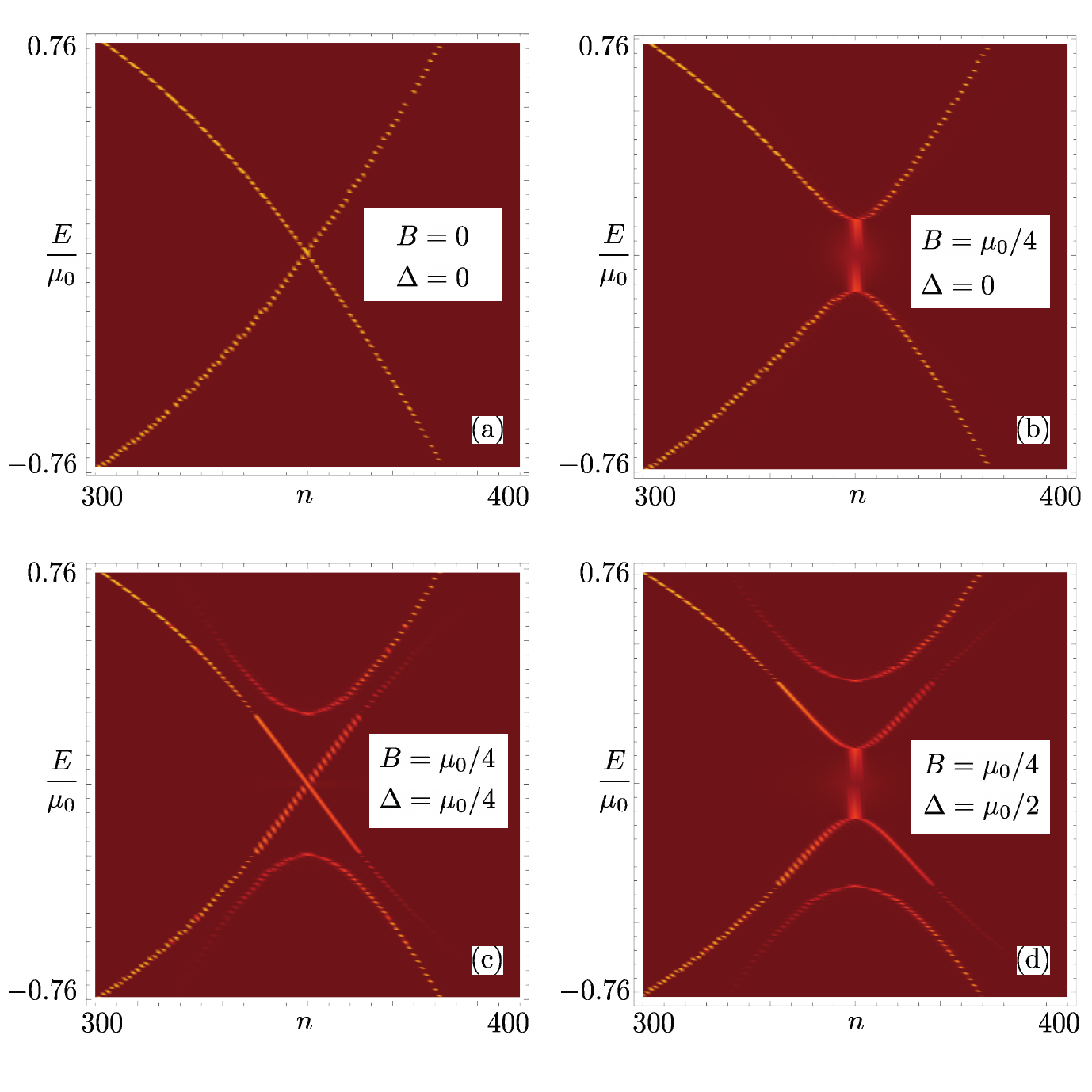}
\caption{Density plot of the spectral weight $w(n,E)$ (see text) as a function of the index $n$ labeling the eigenstates of a particle in a box of 
length $L$ and energy $E$ (units $\mu_0$) for different values of $B$ and 
$\Delta$: $B=\Delta=0$ (a); $B=\mu_0/4$, $\Delta=0$ (b); $B=\Delta=\mu_0/4$ (c); $B=\mu_0/4$, $\Delta=2B$ (d).\label{fig4}}
\end{figure}  
To better understand what is going on, we study the spectral weight $w(n,E)=|c_n[\nu(E)]|^2$, shown as a density plot as a function of the box state index $n$ and the energy $E$ in Fig. \ref{fig4}
\footnote{Note that due to the finite size of the system, the spectrum is 
always discrete and a binning of the energies has been performed to produce the plots.} 
for energies around $E=0$. In general, this tool allows us to get a glimpse of the structure of the energy spectrum of the system. Indeed, as a warm up panel (a) shows the case $B=\Delta=0$: as one can see, the spectral weight is sharply peaked on one box state per energy value and the yellow traces faithfully reproduce the energy spectra of the particle ($E>0$) and hole ($E<0$) sectors of the BdG picture for the Hamiltonian $H=\frac{p^2}{2m}-\mu_0$. Panel (b) shows the situation for $B=\mu_0/4$ and 
$\Delta=0$, when a Tamm state occurs: The energy gap is clearly visible 
and the blurry feature at $E=0$ is the spectral representation of the Tamm state, whose amplitude is smeared across many delocalized box states 
in order to produce a state localized in space. The situation is somewhat 
similar for $\Delta>B$ -- see panel (d) -- again consistent with the picture of a wavefunction with sharp peaks at the system ends. Very peculiar is however the situation for $B=\Delta$, shown in panel (c): at this transition point the spectrum is gapless and exhibits two linear modes around $E=0$. Thus, the zero-energy mode is embedded into a gapless spectrum and yet it still represents a bound state.\\
To summarize our findings so far, even though a gap closing is present at 
the level of the bulk spectrum, bound states can exist for every value of 
$B$ and $\Delta$. However, assessing the nature of such bound states (Tamm-- or Majorana-- like) is not a trivial affair in the context of the full quadratic model which can only be solved numerically. A notable exception is represented by the point $\varphi=0$. In that case, indeed, no Tamm state is present and a bound state is necessarily a Majorana state. However, the general case needs further inspection. In order to tackle this problem and provide an answer, we now proceed developing an analytically solvable low-energy model which is however able to capture all the 
features described so far.
\section{Effective low energy model}
\subsection{Linearization}
To develop an effective, low--energy model valid around the chemical potential we consider a linear dispersion relation instead of the quadratic one. This is meaningful as long as $\mu_0$ is the largest energy scale involved \cite{haldane,ms1,ms2}. In details, we approximate the fermionic operator as \cite{obc1,obc2,obc3,obc4}
\begin{equation}
\psi(x)\simeq e^{ik_Fx}\psi_R(x)+e^{-ik_Fx}\psi_L(x),
\label{eq:dip}
\end{equation}
with $\psi_{R,L}(x)$ fermionic operators for right-- or left--moving electrons. In particular, by using the particle in a box states one can write
\begin{equation}
   \psi_R(x)= \frac{-i}{\sqrt{2L}}\sum_{n=-\infty}^\infty e^{in\pi x/L}d_{n+n_F},
\end{equation}
and $\psi_L(x)=-\psi_R(-x)$ ensuring, together with the $2L$ periodicity 
of the fields, that open boundary conditions $\psi(0)=\psi(L)=0$ are satisfied. Here, the fermionic operators $d_n$ are associated to the $n-$th eigenstate of a particle in a box. Note that, as an approximation, the sum is here extended to {\em all} integers and, consequently, a slight notation abuse has been adopted having defined the operators $d_n$ for all the integer values $n$. This approximation scheme is always performed when discussing the bosonization procedure, and is discussed, for example, in Ref. \cite{giamarchi}.\\
The kinetic energy is then approximated with $K_{L}$
\begin{equation}
K_{L}=v_F \sum_n  \frac{n\pi}{L}d_{n+n_F}=\int_{0}^{L}\left[\psi^\dag_R(x)(-iv_F\partial_x)\psi_R(x)+\psi^\dag_L(x)(iv_F\partial_x)\psi_L(x)\right]dx,
\end{equation}
with $v_F=(\pi n_F)/(mL)$ representing the Fermi velocity.\\
In order to write the full Hamiltonian within the linear approximation, it is useful to introduce the enlarged BdG spinor
\begin{equation}
\Psi_A(x)=(\psi_R(x),\psi_L(x),\psi^\dag_L(x),-\psi^\dag_R(x))^T.
\end{equation}
By imposing that $\mu_0$ is the largest energy scale (we indeed neglect terms proportional to $1/L$ with respect to terms proportional to $k_F$), we finally get the approximate form for the Hamiltonian
\begin{equation}
H\simeq \frac{1}{2}\int_0^L\Psi_A^\dag(x)\mathcal{H}_A(x)\Psi_A(x)dx,
\label{eq:tpp}
\end{equation}
with
\begin{equation}
\mathcal{H}_A(x)=-iv_F\partial_x\tau_3\otimes\tau_3+B\tau_0\otimes(\cos(\phi)\tau_1+\sin(\phi)\tau_2)+\Delta \tau_1\otimes\tau_0.
\end{equation}
\noindent In order to interpret the boundary conditions and the peculiar effects they may have on the bound states, it is useful to think in the following way. Consider a pair of chiral fermions $\chi_R(x)$ and $\chi_L(x)$, independent from each other and organized in the BdG spinor $X(x)=(\chi_R(x),\chi_L(x),\chi^\dag_L(x),-\chi^\dag_R(x))^T$, defined on the whole real axis, and with Hamiltonian 
\begin{equation}
H_{\chi}=\frac{1}{2}\int_{-\infty}^{+\infty} dx~X^\dag (x)\left[\mathcal{H}_A(x)+M\left(x\right)\right]X(x)\,,
\end{equation}
where we have introduced a localized backscattering potential
\begin{equation}
M\left(x\right)=m_0\left[ \delta(x)-\delta (x-L)   \right]\tau_2,
\end{equation}
with $m_0$ the backscattering strength. It can be easily shown that in the limit $m_0/v_F\rightarrow\infty$ the wavefunctions in the region $0<x<L$ become disconnected and acquire the boundary conditions $\chi_L(0)=-\chi_R(0)$, $\chi_L(L)=-\chi_R(-L)$ \cite{timm,tbc1,tbc2}. The Schroedinger problem associated to $H_\chi$, when considered for $0<x<L$ and when $m_0/v_F\rightarrow\infty$ is hence perfectly equivalent to the one related to $H$ {in Eq. \ref{eq:tpp}}.\\
The advantage in resorting to $H_\chi$ is conceptual, since it allows to give a simple interpretation to the results described in the previous section. Indeed, from a physical point of view, $H_\chi$ is equivalent to the Hamiltonian of a quantum spin Hall liquid gapped by both superconductivity, with gap $\Delta$, and a magnetic mass pointing in the direction given by $\phi$ in the presence of two strong backscattering centers with opposite magnetization localized at $x=0,L$.
\subsection{Qualitative interpretation of the results}
\begin{figure}[htbp]	
\widefigure
\includegraphics[width=14 cm]{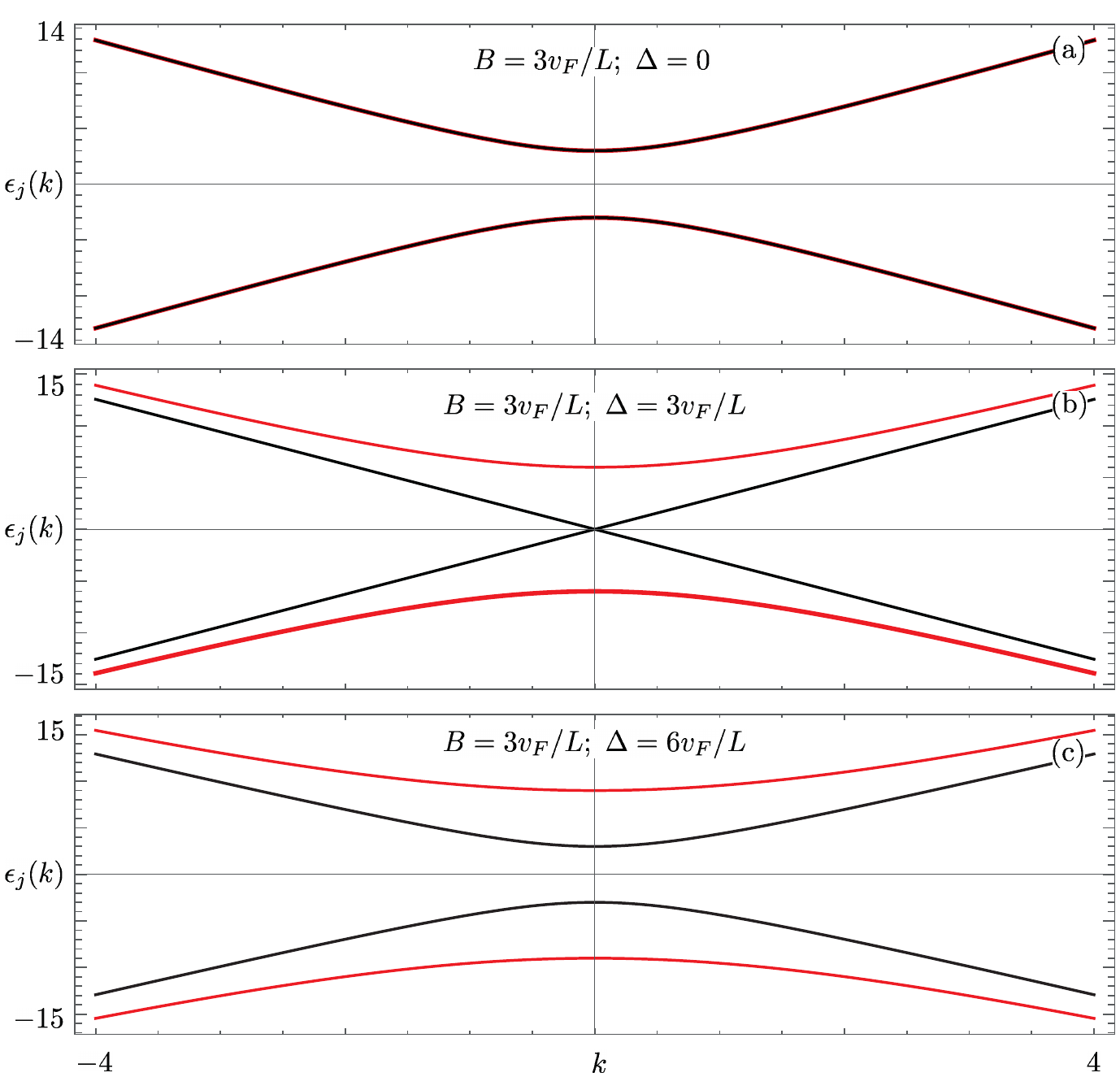}
\caption{Plot of $\epsilon_j(k)$ (units $v_F/L$) as a function of $k$ (units $\pi/L$) for: $B=3v_F/L$, $\Delta=0$ (a); $B=\Delta=3v_F/L$; $B=3v_F/L$, $\Delta=6v_F/L$. In all panels, red lines represent $\epsilon_2(k)$, black lines correspond to $\epsilon_1(k)$.\label{figspecmodel}}
\end{figure}

The first insight into the results comes from the computation of the spectrum as obtained by neglecting the presence of boundaries. If we consider 
the Hamiltonian $H_\infty$ given by
\begin{equation}
    H_\infty=\frac{1}{2}\int_{-\infty}^\infty \Psi_A^\dag(x)\mathcal{H}_A(x)\Psi_A(x)dx,
\end{equation}
where the operators are now being defined, with a slight abuse of notation, on the whole real axis, one can obtain the energy spectrum. The spectrum is promptly found to be $\epsilon_{j}(k)=\pm\sqrt{v_F^2k^2+\Delta^2+B^2+(-1)^j 2\Delta B }$ with $j\in\{1,2\}$ and is shown in Fig. \ref{figspecmodel}.
As can be seen, the spectrum is always gapped for all $B>0,\,\Delta>0$ except for $B=\Delta $ -- see panel (b) -- where $\epsilon_{1}(k)$ is gapless. The phase space point $B=\Delta$ marks a 
quantum phase transition where the gap changes from magnetic ($B>\Delta $) to superconducting ($B<\Delta $). It is however crucial to notice the presence of a secondary gap of magnitude $B+\Delta $ in $\epsilon_{2}(k)$ which is always present for every $B>0,\,\Delta>0$. Notice also the qualitative similarity between these spectra and the spectral features found in 
the quadratic model substantiating the validity of the linear approximation at low energies, compared to $\mu_0$.\\

\noindent Turning to the inspection of bound states it is crucial to observe that the addition of a confining term proportional to $m_0$ produces a magnetic gap in the spectrum. It is well known \cite{alicea} that in the superconducting dominated regime ($B<\Delta $) Majorana bound states appear at the boundaries of the system. However, in the opposite case, when $B>\Delta $ the situation is more subtle, as this model would implement the physics of so called Goldstone-Wilczek fractional solitons \cite{gw6}. Since this regime is adiabatically connected to $\Delta =0$, in order to build some physical intuition one can refer to the case in which superconductivity is absent. The system is then equivalent to a quantum spin Hall liquid, in the absence of superconductivity, gapped by two magnets of equal strength at $x=0,\,L$ and characterized by different magnetization angles: it hence hosts bound states. Assuming the decay length of these bound states to be shorter than the system size $L$, the two bound states are essentially independent, and thus it is sufficient to concentrate on a single boundary to capture the relevant physics.
\subsection{Linear model with one boundary and $\Delta = 0$}
We thus inspect a simpler linear model in the regime $\Delta=0$, $L\rightarrow\infty$ considering the bound state localized around $x=0$. Its wavefunction $\psi^{}_m(x)$, that since superconductivity is now absent ($\Delta=0$) has only the two right-- and left--moving components, can be easily obtained and reads 
\begin{equation}
 \psi_m^{}(x) =\sqrt{\frac{m}{2v_F}{\sin\left(\phi\right)}}\left(-1 ,1 \right)^T  e^{\frac{m}{v_F}{\sin\left(\phi\right)}x}.
\label{eq:lin}\end{equation}
The solution is only acceptable (normalizable) for $\pi<\phi<2\pi$. The corresponding energy is $\epsilon_0^{}= -m\cos(\phi)$. Note that for $\phi=3\pi/2$ the state, mimiking the Tamm state of the quadratic model, is at zero energy.
The scenario just derived perfectly agrees with the behavior of the left 
boundary of the quadratic model, although in that case the wavefunction in Eq. \ref{eq:lin} by means of Eq. \ref{eq:dip} has one component only, and is rebuilt from the condition in Eq. \ref{eq:dip}. Moreover, in the quadratic model, the unphysical particle-hole symmetry of the Bogoliubov-de Gennes equation is present while here the redundancy\cite{franchini,xy} is not needed since superconductivity is not present. We hence have a single bound state instead of two at opposite energies.\\
The addition of a second physical boundary at $x=L$ produces effects that can be now easily understood by leveraging on the single boundary case 
just discussed.
Indeed, the bound state at the boundary at $x=L$ is present when $\sin(\phi)>0$ -- that is for $0<\phi<\pi$ -- when the bound state is absent at 
$x=0$. Such exponentially increasing solution is admissible due to the added barrier at $x=L$ -- which prevents the divergence of the wavefunction -- and its decay through the dot. This again is perfectly compatible with the results of the quadratic model discussed in Sec. 2.\\
A final comment for this section is the behavior at $\phi=0,\,\pi$. For those values of the angle, the energy of the bound state reaches the bulk bands and the bound state becomes delocalized, and hence, effectively, disappears.\\
In the following we will concentrate on the case in which the bound state is maximally localized around $x=0$, that is for $\phi=3\pi/2$.
\subsection{The wavefunction of the bound state for $\varphi=3\pi/2$}
A further interesting behavior of the bound state found within the quadratic model is the fact that, when present, it does not become completely delocalized at the quantum phase transition $B=\Delta$. To confirm that this fact occurs also within the linearized model, we here report the BdG expression of the wavefunction $\psi_0(x)$ of the bound state characterizing the linearized model in Eq. \ref{eq:tpp}, together with the condition in Eq. \ref{eq:dip}, for $\phi=3\pi/2$ and arbitrary $B>0,\,\Delta>0$. We find
\begin{equation}
\psi_0(x)=(u(x),v(x))^T\,,
\end{equation}
where
\begin{align*}
    u(x)=&e^{ik_Fx}\chi_0(x)-e^{-ik_Fx}\chi_0(-x),\\
    v(x)=&e^{ik_Fx}\xi_0(x)-e^{-ik_Fx}\xi_0(-x),\\
    \chi_0(x)=&\bigg[\bigg(\frac{\Delta-B}{\Delta+B}\frac{\sinh{[L(\Delta+B)/v_F]}}{\sinh{[L(\Delta-B)/v_F]}}\frac{\Delta+B}{4v_F}\frac{e^{-2\Delta L/v_F}}{1-e^{-2(\Delta+B)L/v_F}}\bigg)^{1/2}e^{(\Delta-B)x/v_F}+\\
    &+\bigg(\frac{\Delta+B}{4v_F}\frac{1}{1-e^{-2(\Delta+B)L/v_F}}\bigg)^{1/2}e^{-(\Delta+B)x/v_F}\bigg]\Theta(x)+\\
    &+\bigg[\bigg(\frac{\Delta-B}{\Delta+B}\frac{\sinh{[L(\Delta+B)/v_F]}}{\sinh{[L(\Delta-B)/v_F]}}\frac{\Delta+B}{4v_F}\frac{e^{-2\Delta L/v_F}}{1-e^{-2(\Delta+B)L/v_F}}\bigg)^{1/2}e^{-(\Delta-B)x/v_F}+\\
    &+\bigg(\frac{\Delta+B}{4v_F}\frac{1}{1-e^{-2(\Delta+B)L/v_F}}\bigg)^{1/2}e^{(\Delta+B)x/v_F}\bigg]\Theta(-x)\\
    \xi_0(x)=&\bigg[\bigg(\frac{\Delta-B}{\Delta+B}\frac{\sinh{[L(\Delta+B)/v_F]}}{\sinh{[L(\Delta-B)/v_F]}}\frac{\Delta+B}{4v_F}\frac{e^{-2\Delta L/v_F}}{1-e^{-2(\Delta+B)L/v_F}}\bigg)^{1/2}e^{(\Delta-B)x/v_F}+\\
    &-\bigg(\frac{\Delta+B}{4v_F}\frac{1}{1-e^{-2(\Delta+B)L/v_F}}\bigg)^{1/2}e^{-(\Delta+B)x/v_F}\bigg]\Theta(x)+\\
    &+\bigg[\bigg(\frac{\Delta-B}{\Delta+B}\frac{\sinh{[L(\Delta+B)/v_F]}}{\sinh{[L(\Delta-B)/v_F]}}\frac{\Delta+B}{4v_F}\frac{e^{-2\Delta L/v_F}}{1-e^{-2(\Delta+B)L/v_F}}\bigg)^{1/2}e^{-(\Delta-B)x/v_F}+\\
    &-\bigg(\frac{\Delta+B}{4v_F}\frac{1}{1-e^{-2(\Delta+B)L/v_F}}\bigg)^{1/2}e^{(\Delta+B)x/v_F}\bigg]\Theta(-x).
\end{align*}
Both the particle and the hole components $u(x)$ and $v(x)$ are characterized by a profile that is peaked around the edge {\em even at the transition point $B=\Delta$}.
\begin{figure}[htbp]	
\widefigure
\includegraphics[width=14 cm]{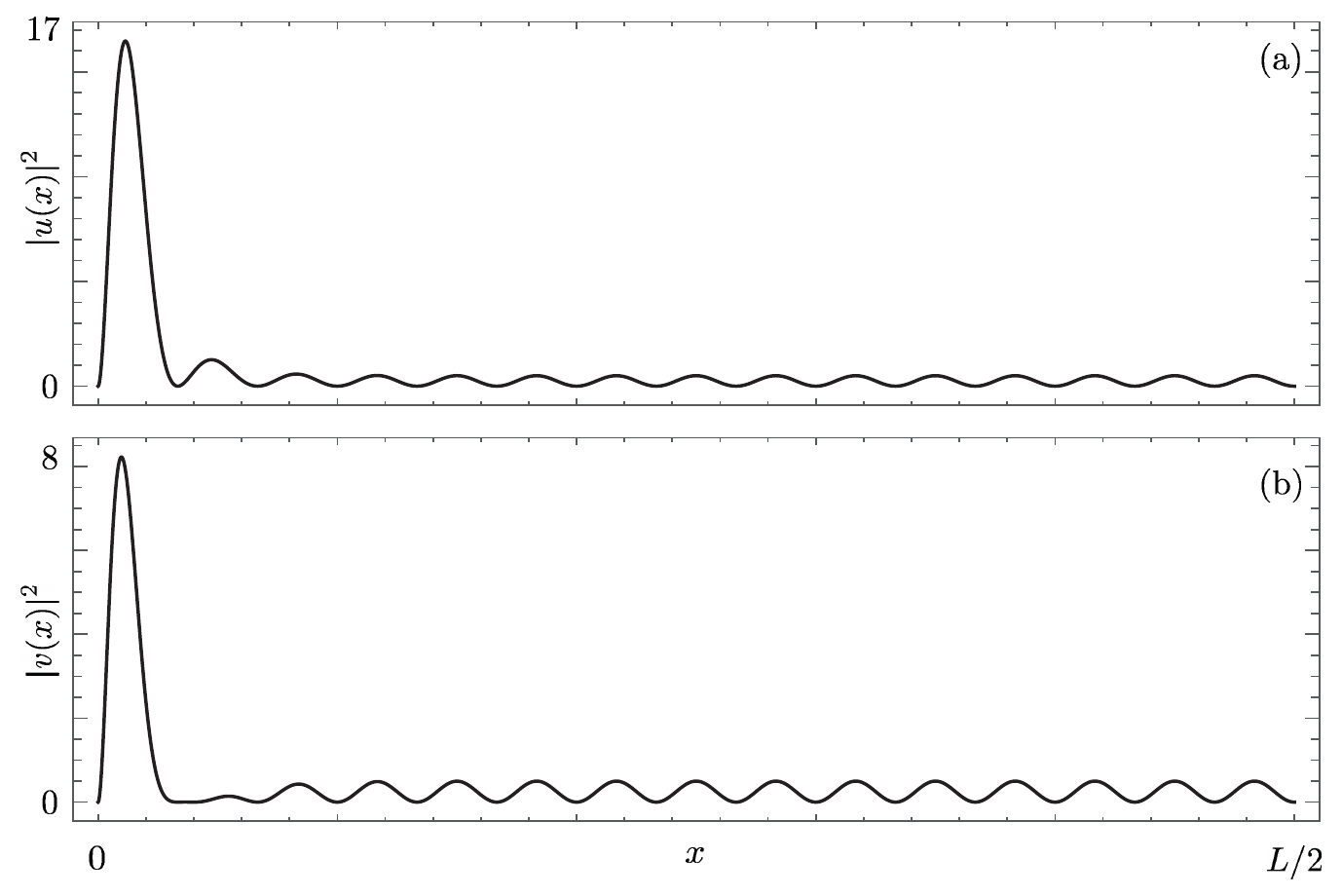}
\caption{Plot of the probability amplitude of particle and hole components of the zero--mode wavefunction (units $L^{-1}$) as a function of $x$: $|u(x)|^2$ (a); $|v(x)|^2$ (b). Here, $B=\Delta=30v_F/L$ and $k_F=30\pi/L$.\label{figwfanal}}
\end{figure}  
This case is shown in Fig. \ref{figwfanal}, where a plot of $|u(x)|^2$ and $|v(x)|^2$ is reported for $B=\Delta$: one can clearly notice that at the transition point, although the wavefunction is sharply localized at $x=0$, it does not completely decay exponentially to zero. Also this fact is in excellent agreement with the results obtained in the full quadratic model. As apparent from the analytical form of the wavefunction, this fact is due to the presence of the $B+\Delta$ energy scale.\\
A natural question hence arises: are the Majorana zero modes, in the superconducting dominated phase $B<\Delta$, influenced by the persistence of the Tamm state across the quantum phase transition? 
\section{Majorana polarization}
The Majorana polarization $P_M$ \cite{mp1} is a quantity that has been proposed in order to discriminate between Majorana fermions and other types of Andreev bound states. We now evaluate the Majorana polarization associated to the state $\psi_0(x)$, defined as
\begin{equation}
P_M= \frac{N_L}{D_L},
\end{equation}
where
\begin{equation}
 N_L = \int_0^{L / 2} 2 u (x) v (x) d x ;\,\,\,\,\,\, D_L = \int_0^{L 
/ 2} |u (x)
   |^2 + | v (x) |^2 d x .
\end{equation}
This absolute value of $P_M$ is close to one (in the large gap limit) if the bound state is a Majorana bound state, while it reaches different values for regular Andreev bound states, which allows to assess the character of $\psi_0(x)$. In the 
case under examination we find
\begin{equation}
P_M=\frac{\left\{L^2k_F^2 + \left[\frac{L(B - \Delta)}{v_F}\right]^2\right\} \left[1 + e^{\frac{L(B - \Delta)}{v_F}}\right]
   - \left\{L^2k_F^2 + \left[\frac{L(B + \Delta)}{v_F}\right]^2\right\} \left[1 + e^{- \frac{L(B - \Delta)}{v_F}}\right]}{\left\{L^2k_F^2 + \left[\frac{L(B - \Delta)}{v_F}\right]^2\right\} \left[1 + e^{\frac{L(B - \Delta)}{v_F}}\right] + \left\{L^2k_F^2 +  \left[\frac{L(B + \Delta)}{v_F}\right]^2\right\} \left[1 + e^{-
   \frac{L(B + \Delta)}{v_F})}\right]} .
\end{equation}
Looking at the previous expressions, two limiting cases can be extracted: 
for $B=0$ one finds $P_M=\tanh\left(\frac{L\Delta}{v_F}\right)$, with 
$P_M \to 1$ when $\Delta\to\infty$. For $\Delta\ll B$ (and $\mu_0>> B$) on the other hand we find $P_M\rightarrow -\frac{2B\Delta}{\mu_0^2}$ and thus $P_M\to0^{-}$. From the point of view of the Majorana polarization we 
hence have that in the phase dominated by the magnetic field, the Majorana polarization is small, while it approaches, in modulus, the unity in the large gap limit. A transition takes place for $B\simeq\Delta$. However, 
such a transition does not happen at the exact value of the gap closing, even for $L\rightarrow\infty$, that is $B=\Delta$. This behaviour of $P_M$ shows that, despite the fact that the Majorana phase indeed hosts polarized Majorana fermions, a reminiscence of the Tamm state is however present -- and visible -- for $B=\Delta$.
\begin{figure}[H]	
\widefigure
\includegraphics[width=14 cm]{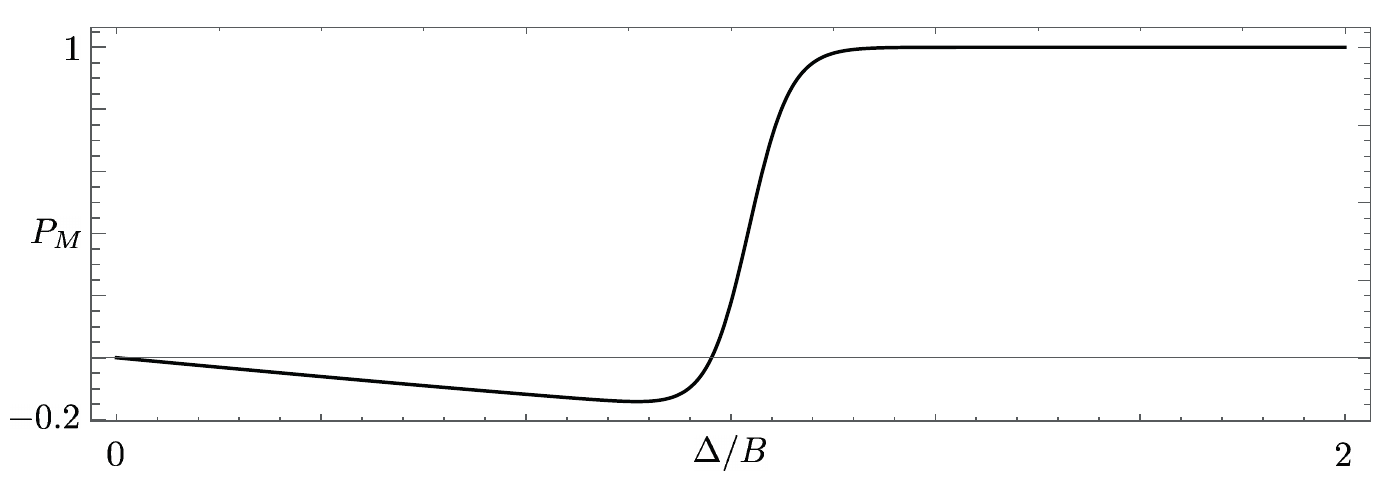}
\caption{Plot of the Majorana polarization $P_M$ as a function of $\Delta/B$ for $k_F=30\pi/L$ and $B=30v_F/L$\label{Fig:MP}.}
\end{figure}  
The Majorana polarization is shown in Fig. \ref{Fig:MP}.
\section{Conclusions}
In this work we have characterized a simple model for a finite size, one dimensional topological superconductor in the presence of a competing normal gapping mechanism arising from a position dependent potential. The first part of the article deals with the numerical inspection of the model. We find that the model has two phases, one characterized by Majorana fermions at both the ends, and the other by at most a single Tamm bound state localized at one side. The energy of the bound state, when present, strongly depends on the phase of the position dependent potential. Surprisingly, we find that the wavefunction of the bound state remains strongly peaked close to the edges even at the gap closing point separating the two phases. In the second part of the article, we develop an exactly solvable model for interpreting the results of the first part. Finally, in the last section we evaluate the Majorana polarization for a fully developed bound state of the model and show that, despite the fact that the Tamm state is not completely delocalized at the transition point, deep in the superconducting phase the Majorana bound states are not strongly affected by the position--dependent potential.

\authorcontributions{Conceptualization, M. C., N. T. Z. and F. C.; methodology, L. V.; software, M. C. and F. C.;
validation, F. C., M. C. and L. V.; formal analysis, L. V. and N. T. Z.; writing original draft preparation, N. T. Z., F. C.; all authors
have read and agreed to the published version of the manuscript.}

\funding{This research received no external funding.}

\acknowledgments{We acknowledge fruitful discussions with Simone Traverso and Maura Sassetti.}

\conflictsofinterest{The authors declare no conflict of interest.} 
\end{paracol}


\begin{thebibliography}{999}
\bibitem{das1}Nayak, C.; Simon, S. H.; Stern, A.; Freedman, M.;
Das Sarma, S. Non-Abelian Anyons and Topological Quantum Computation. {\em Rev. Mod. Phys.} {\bf2008}, {\em 80}, 1083.
\bibitem{para1}Read, N.; Rezayi, E. Beyond paired quantum hall states:
Parafermions and incompressible states in the first excited landau level. {\em Phys. Rev. B} {\bf 1999}, {\em59}, 8084.
\bibitem{teo1}
Blasi, A.; Braggio, A.; Carrega, M.; Ferraro, D.; Maggiore, N.; Magnoli, N. Non-Abelian BF theory for 2+1 dimensional topological states of matter. {\em New J. Phys.} {\bf 2012}, {\em 14}, 013060.
\bibitem{para2}Zhang, F.; Kane, C. L. {\em Phys. Rev. Lett.} {\bf 2014}, {\em 113}, 036401.
\bibitem{para3}Orth, C. P.; Tiwari, R. P.; Meng, T.; Schmidt, T. L. Non-Abelian parafermions in time-reversal-invariant interacting helical systems. {\em Phys.
Rev. B} {\bf 2015}, {\em 91}, 081406(R).
\bibitem{para4}Klinovaja, J.; Loss, D. Fractional charge and spin states in topological insulator constrictions. {\em Phys. Rev. B} {\bf 2015}, {\em 92}, 121410(R).
\bibitem{para5}Fleckenstein, C.; Traverso Ziani, N.; Trauzettel, B. $Z_4$ parafermions in Weakly Interacting Superconducting Constrictions at the Helical Edge of Quantum Spin Hall Insulators. {\em Phys. Rev. Lett.} \textbf{2020}, {\em 122}, 066801.
\bibitem{teo2}
Rossini, D.; Carrega, M.; Calvanese Strinati, M.; Mazza, L. Anyonic tight-binding models of parafermions and of fractionalized fermions. {\em Phys. Rev. B} {\bf 2019}, {\em 99}, 085113.
\bibitem{teo3}
Carrega, M.; Principi, A.; Vera-Marun, I. Tunneling spectroscopy as a probe of fractionalization in 2D magnetic heterostructures. {\em Phys. Rev. B} {\bf 2020}, {\em 102}, 085412.
\bibitem{df1}Ronetti, F.; Carrega, M.; Ferraro, D.; Rech, J.; Jonckheere, T.; Martin, T.; Sassetti M. Polarized heat current generated by quantum pumping in two-dimensional topological insulators.
{\em Phys. Rev. B} {\bf 2017}, {\em 95}, 115412.
\bibitem{addd1}Schiller, N.; Cornfeld, E.; Berg, E.; Oreg, Y. Predicted signatures of topological superconductivity and parafermion zero modes in fractional quantum Hall edges. {\em Phys. Rev. Research} {\bf 2020}, {\em 2}, 023296.
\bibitem{addd2}Michelsen, A. B.; Schmidt, T. L.: Idrisov, E. G. Current correlations of Cooper-pair tunneling into a quantum Hall system. {\em Phys. Rev. B} {\bf 2020}, {\em 102}, 125402.
\bibitem{addd3}Zhao, L.; Arnault, E. G.; Bondarev, A.; Seredinski, A.;
Larson, T. F. Q.: Draelos, A. W.; Li, H.; Watanabe, K.; Taniguchi, T.;
Amet, F.; Baranger, H. U.; Finkelstein, G. Interference of chiral
Andreev edge states, {\em Nat. Phys.} {\bf 2020}, {\em 16}, 862.
\bibitem{suphall1}Amet, F.; Ke, C. T.; Borzenets, I. V.; Wang, J.; Watanabe, K.;
Taniguchi, T.; Deacon, R. S.; Yamamoto, M.; Bomze, Y.;
Tarucha, S.; Finkelstein, G. Supercurrent in the quantum Hall regime. {\em Science} {\bf 2016}, {\em 352}, 966.
\bibitem{suphall2}Lee, G.-H.; Huang, K.-F.; Efetov, D. K.; Wei, D. S.; Hart, S.; Taniguchi, T.; Watanabe, K.; Yacoby, A.; Kim, P. Inducing
Superconducting Correlation in Quantum Hall Edge States. {\em Nat. Phys.} {\bf 2017}, {\em 13}, 693.
\bibitem{suphall3}G\"ul, O.; Ronen, Y.; Lee, S. Y.; Shapourian, H.; Zauberman, J.; Lee, Y. H.; Watanabe, K.; Taniguchi, T.; Vishwanath, A.; Yacoby, A.; Kim P. Induced superconductivity in the fractional quantum Hall edge. arXiv:2009.07836.
\bibitem{majo1}Mourik, V.; Zuo, K.; Frolov, S. M.; Plissard, S. R.;  Bakkers, E. P. A. M.; Kouwenhoven, L. P. Signatures of Majorana fermions in hybrid superconductor-semiconductor nanowire devices. {\em Science} {\bf 2012}, {\em 336}, 1003.
\bibitem{majo2}Lutchyn, R. M.; Sau, J. D.; Das Sarma, S. Majorana Fermions and a Topological Phase Transition in Semiconductor-Superconductor Heterostructures. {\em Phys. Rev. Lett.}
{\bf 2010} {\em 105}, 077001.
\bibitem{majo3}Oreg, Y.; Refael, G.; von Oppen, F. Helical Liquids and Majorana Bound States in Quantum Wires. {\em Phys. Rev. Lett.} {\bf 2010}, {\em 105}, 177002.
\bibitem{fu1}Fu, L.; Kane, C. L. Superconducting proximity effect and Majorana fermions at the surface of a topological insulator. {\em Phys. Rev. Lett.} {\bf 2008}, {\em 100}, 096407.
\bibitem{fu2}Fu, L.; Kane, C. L. Josephson current and noise at a superconductor/quantum-spin-Hall-insulator/superconductor junction. {\em Phys. Rev. B} {\bf 2009}, {\em 79}, 161408.
\bibitem{cb1}Akhmerov, A. R.; Nilsson, J.; Beenakker, C. W. J. Electrically detected interferometry of Majorana fermions in a topological insulator.
{\em Phys. Rev. Lett.} {\bf 2009}, {\em 102}, 216404.
\bibitem{qpc}Strunz, J.; Wiedenmann, J.; Fleckenstein, C.; Lunczer, L.; Beugeling, W.; Müller, V. L.; Shekhar, P.; Traverso Ziani, N.; Shamim, S.;
Kleinlein, J.; et al. Interacting topological edge channels. {\em Nat. Phys.} {\bf 2020}, {\em 16}, 83.
\bibitem{qpc1}Li, J.; Pan, W.; Bernevig, B. A.; Lutchyn, R. M. Detection of Majorana Kramers Pairs Using a Quantum Point Contact. {\em Phys. Rev.
Lett.} {\bf 2016}, 117, 046804.
\bibitem{qpc2}Fleckenstein, C.; Traverso Ziani, N.; Calzona, A.; Sassetti, M.; Trauzettel B. Formation and detection of Majorana modes in quantum spin Hall trenches. {\em Phys. Rev. B} {\bf 2021}, {\em 103}, 125303.
\bibitem{fc}Nadj-Perge, S.; Drozdov, I. K.; Li, J.; Chen, H.; Jeon, S.; Seo, J.; MacDonald, A. H.; Bernevig, B. A.; Yazdani A. Observation of Majorana fermions in ferromagnetic atomic chains on a superconductor. {\em Science} {\bf 2014}, {\em 6209}, 602.
\bibitem{pjj}Ren, H.; Pientka, F.; Hart, S.;  Pierce, A. T.; Kosowsky, M.; Lunczer, L.; Schlereth, R.; Scharf, B.; Hankiewicz, E. M.; Molenkamp, L. W.;  Halperin, B. A.; Yacoby A. Topological superconductivity in a phase-controlled Josephson junction. {\em Nature} {\bf 2019}, {\em 569}, 93.
\bibitem{teo5}
Guiducci, S.; Carrega, M.; Biasiol, G.; Sorba, L.; Beltram, F.; Heun, S. Toward Quantum Hall Effect in a Josephson Junction. {\em Phys. Status Solidi RRL} {\bf 2019}, {\em 13}, 1800222.
\bibitem{teo6}
Guiducci, S.; Carrega, M.; Taddei, F.; Biasiol, G.; Courtois, H.; Beltram, F.; Heun, S. Full electrostatic control of quantum interference in an extended trenched Josephson junction. {\em Phys. Rev. B} {\bf 2019}, {\em 99}, 235419.
\bibitem{mf11}Krogstrup, P.; Ziino, N. L. B.; Chang, W.; Albrecht, S. M.; Madsen, M., H.; Johnson, E.; Nygard, J.; Marcus, C., M.; Jespersen, T. S. Epitaxy of semiconductor–superconductor nanowires. {\em Nat. Mater.} {\bf 2015}, {\em 14}, 400.
\bibitem{mf12}Deng, M. T.; Vaitiekenas, S.; Hansen, E. B.; Danon, J.; Leijnse, M.; Flensberg, K.; Nygard, J.; Krogstrup, P.; Marcus, C. M. Majorana bound state in a coupled quantum-dot hybrid-nanowire system. {\em Science} {\bf 2016}, {\em 354}, 1557.
\bibitem{aguado}Aguado, R. Majorana quasiparticles in condensed matter. {\em La Rivista del Nuovo Cimento} {\bf 2017}, {\em 40}, 523.
\bibitem{abs0}Asano, Y.; Tanaka, Y.; Kashiwaya, S. Phenomenological
theory of zero-energy Andreev resonant states. {\em Phys. Rev. B} {\bf 2004}, {\em 69}, 134501.
\bibitem{abs1}Tanaka, Y.; Kashiwaya, S.; Yokoyama, T. Theory of
enhanced proximity effect by midgap Andreev resonant state in
diffusive normal-metal/triplet superconductor junctions. {\em Phys.
Rev. B} {\bf 2005}, {\em 71}, 094513.
\bibitem{abs2}Golubov, A. A.; Brinkman, A.; Tanaka, Y.; Mazin, I. I.;
Dolgov, O. V. Andreev Spectra and Subgap Bound States
in Multiband Superconductors. {\em Phys. Rev. Lett.} {\bf 2009} {\em 103}, 077003.
\bibitem{abs3}Tanaka, Y.; Mizuno, Y.; Yokoyama, T.; Yada, K.; Sato, M.
Anomalous Andreev Bound State in Noncentrosymmetric Superconductors, {\em Phys. Rev. Lett.} {\bf 2012}, {\em 105}, 097002.
\bibitem{abs4}Liu, J.; Potter, A. C.; Law, K. T.; Lee, P. A. Zero-Bias
Peaks in the Tunneling Conductance of Spin-Orbit-Coupled
Superconducting Wires with and without Majorana End-States.
{\em Phys. Rev. Lett.} {\bf 2012}, {\em 109}, 267002.
\bibitem{abs5}Kells, G.; Meidan, D.; Brouwer, P. W. Near-zero-energy end
states in topologically trivial spin-orbit coupled superconducting nanowires with a smooth confinement, {\em Phys. Rev. B} {\bf 2012}, {\em 86},
100503(R).
\bibitem{abs6}Roy, D.; Bondyopadhaya, N.; Tewari, S. Topologically
trivial zero-bias conductance peak in semiconductor Majorana
wires from boundary effects, {\em Phys. Rev. B} {\bf 2013}, {\em 88}, 020502(R).
\bibitem{abs7}Stanescu, T. D.; Tewari, S. Disentangling Majorana
fermions from topologically trivial low-energy states in semiconductor Majorana wires, {\em Phys. Rev. B} {\bf 2013} {\em 87}, 140504(R).
\bibitem{abs8}Cayao, J.; Prada, E.; San-Jose, P.; R. Aguado, R. SNS junctions
in nanowires with spin-orbit coupling: Role of confinement
and helicity on the subgap spectrum. {\em Phys. Rev. B} {\bf 2015}, {\em 91}, 024514.
\bibitem{abs9}San-Jose, P.; Cayao, J.; Prada, E.; Aguado, R. Majorana
bound states from exceptional points in non-topological superconductors. {\em Sci. Rep.} {\bf 2016}, {\em 6}, 21427.
\bibitem{abs10}Liu, C.-X.; Sau, J. D.; Stanescu, T. D.; Das Sarma, S.
Andreev bound states versus Majorana bound states in quantum
dot-nanowire-superconductor hybrid structures: Trivial versus
topological zero-bias conductance peaks. {\em Phys. Rev. B} {\bf 2017}, {\em 96}, 075161.
\bibitem{abs11} Liu, C.-X.; Sau, J. D.; and Das Sarma, S. Distinguishing topological Majorana bound states from trivial Andreev bound states:
Proposed tests through differential tunneling conductance spectroscopy. {\em Phys. Rev. B} {\bf 2018} {\em 97}, 214502.
\bibitem{abs12}Moore, C.; Stanescu, S. T.; Tewari, S. Two-terminal
charge tunneling: Disentangling Majorana zero modes from
partially separated Andreev bound states in semiconductorsuperconductor heterostructures. {\em Phys. Rev. B} {\bf 2018}, {\em 97}, 165302.
\bibitem{abs13}Moore, C.; Zeng, C.; Stanescu,  T. D.; Tewari, S.
Quantized zero-bias conductance plateau in semiconductorsuperconductor heterostructures without topological Majorana zero modes. {\em Phys. Rev. B} {\bf 2018}, {\em 98}, 155314.
\bibitem{abs14}Fleckenstein, C.; Dominguez, F.; Traverso Ziani, N.; 
Trauzettel, B. Decaying spectral oscillations in a Majorana wire
with finite coherence length. {\em Phys. Rev. B} {\bf 2018}, {\em 97}, 155425.
\bibitem{abs15}Awoga, O. A.; Cayao, J.; Black-Schaffer, A. M. Supercurrent Detection of Topologically Trivial Zero-Energy States in
Nanowire Junctions. {\em Phys. Rev. Lett.} {\bf 2019}, {\em 123}, 117001.
\bibitem{abs16}Marra, P.; Nitta, M. Topologically nontrivial Andreev bound states. {\em Phys. Rev. B} {\bf 2019}, {\em 100}, 220502(R).
\bibitem{abs17}Cayao, J.; Black-Schaffer, A. M. Distinguishing trivial and topological zero energy states in long nanowire junctions. arXiv:2011.10411.
\bibitem{gbu1}
Pan, H.; Das Sarma, S. Physical mechanisms for zero-bias conductance peaks in Majorana nanowires. {\em Phys. Rev. Research.} {\bf 2020}, {\em 2}, 013377.
\bibitem{gbu2}Huang, Y.; Pan, H.; Liu, C.-X.; Sau, J. D.;Stanescu, T. D.; Das Sarma, S. Metamorphosis of Andreev bound states into Majorana bound states in pristine nanowires. {\em Phys. Rev. Lett.} {\bf 2018}, {\em 98}, 144511.
\bibitem{tamm1}Gangadharaiah, S.; Trifunovic, L.; Loss, D. Localized End States in Density Modulated Quantum Wires and Rings. {\em Phys. Rev. Lett.} {\bf 2012} {\em 108}, 136803.
\bibitem{tamm2}Traverso Ziani, N.; Fleckenstein, C.; Vigliotti, L.; Trauzettel, B.; Sassetti, M. From fractional solitons to Majorana fermions in a paradigmatic model of topological superconductivity. {\em Phys. Rev. B} {\bf 2020}, {\em 101}, 195303.
\bibitem{tamm3}Henriques, J. C. G.; Rappoport, T. G.; Bludov, Y. V.; Vasilevskiy, M. I.; Peres, N. M. R. Topological photonic Tamm states and the Su-Schrieffer-Heeger model. {\em Phys. Rev. A} {\bf 2020}, {\em 101}, 043811.
\bibitem{gw1}Jackiw, R.; Rebbi, C. Solitons with fermion number ½. {\em Phys. Rev. D} {\bf{1976}}, {\em 13}, 3398.
\bibitem{gw11}Goldstone, J.; Wilczek, F. Fractional Quantum Numbers on Solitons. {\em Phys. Rev. Lett.} {\bf 1981}, {\em 47}, 986.
\bibitem{gw2}Kivelson, S.; Schrieffer, J. R. Wannier functions in one-dimensional disordered systems: Application to fractionally charged solitons. {\em Phys. Rev. B} {\bf 1982}, {\em 25}, 6447.
\bibitem{gw3}Qi, X.-L.; Hughes, T. L.; Zhang, S.-C. Fractional charge and quantized current in the quantum spin Hall state. {\em Nat. Phys.} {\bf 2008}, {\em 4}, 273.
\bibitem{gw4}V\"ayrynen J. I.; Ojanen, T. Chiral Topological Phases and Fractional Domain Wall Excitations in One-Dimensional Chains and Wires. {\em Phys. Rev. Lett.} {\bf 2011}, {\em 107}, 166804.
\bibitem{gw5}Klinovaja, J.; Stano, P.; Loss, D. Transition from Fractional to Majorana Fermions in Rashba Nanowires. {\em Phys. Rev. Lett.} {\bf 2012}, {\em 109}, 236801.
\bibitem{gw6}Fleckenstein, C.; Traverso Ziani, N.; Trauzettel, B. Chiral anomaly in real space from stable fractional charges at the edge of a quantum spin Hall insulator. {\em Phys.
Rev. B} {\bf 2016}, {\em 94}, 241406(R).
\bibitem{gw8}
Gresta, D.; Blasi, G.; Taddei, F.; Carrega, M.; Braggio, A.; Arrachea, L. Signatures of Jackiw-Rebbi resonance in the thermal conductance of topological Josephson junctions with magnetic islands. {\em Phys. Rev. B} {\bf 2021}, {\em 103}, 075439.
\bibitem{mp1}Sticlet, D.; Bena, C.; Simon, P. Spin and Majorana Polarization in Topological Superconducting Wires. {\em Phys. Rev. Lett.} {\bf 2012}, {\em 108}, 096802.
\bibitem{mp2} Sedlmayr, N.; Bena, C. Visualising Majorana bound states in 1D and 2D using the generalized Majorana polarization. {\em Phys. Rev. B} {\bf 2015}, {\em 92}, 115115.
\bibitem{mp3}Bena, C. Testing the formation of Majorana states using Majorana polarization. {\em Comptes Rendus Physique} {\bf 2017}, {\em 18}, 349.
\bibitem{bernevig}Bernevig, B. A.; Hughes, T. L. Topological insulators and
topological superconductors. (Princeton University Press 2013, Princeton, NJ).
\bibitem{joha}Malard, M.; Japaridze, G. I.; Johannesson, H. Synthesizing Majorana zero-energy modes in a periodically gated quantum wire.
{\em Phys. Rev. B} {\bf 2016}, {\em 94}, 115128.
\bibitem{peierls}Peierls, R. A. Zur Theorie der elektrischen und thermischen Leitfähigkeit von Metallen. {\em Ann. Phys.} {\bf 1930}, {\em 4}, 121.
\bibitem{haldane} Haldane, F. D. M. Effective harmonic-fluid approach to low-energy properties of one-dimensional quantum fluids. {\em Phys. Rev. Lett.} {\bf 1981}, {\em 47}, 1840.
\bibitem{ms1}Cuniberti, G.; Sassetti, M.; Kramer, B.
Transport and elementary excitations of a Luttinger liquid.
{\em J. Phys. Condens. Matter} {\bf 1996}, {\em 8}, L21.
\bibitem{ms2}Guinea, F.; Santos, G. G.; Sassetti, M.; Ueda, M.
Asymptotic tunneling conductance in Luttinger liquids.
{\em Europhys. Lett} {\bf 80}, {\em 30}, 561.
\bibitem{obc1}Fabrizio, M.; Gogolin, A. O. Interacting one-dimensional electron gas with open boundaries. {\em Phys. Rev. B} {\bf 1995}, {\em 51}, 17827.
\bibitem{obc2} Traverso Ziani, N.; Cavaliere, F.; Sassetti, M. Signatures of Wigner correlations in the conductance of a one-dimensional quantum
dot coupled to an AFM tip. {\em Phys. Rev. B} {\bf 2012}, {\em 86}, 125451.
\bibitem{obc3}Traverso Ziani, N.; Cavaliere, F.; Sassetti, M. Theory of the STM detection of Wigner molecules in spin-incoherent CNTs. {\em Europhys.
Lett.} {\bf 2013}, {\em 102}, 47006.
\bibitem{obc4} Porta, S.; Gambetta, F. M.; Traverso Ziani, N.; Kennes, D. M.; Sassetti, M.; Cavaliere, F. Nonmonotonic response and light-cone
freezing in fermionic systems under quantum quenches from gapless to gapped or partially gapped states. {\em Phys. Rev. B} {\bf 2018}, {\em 97},
035433.
\bibitem{timm}Timm, C. Transport through a quantum spin Hall quantum dot.
{\em Phys. Rev. B} {\bf 2012}, {\em 86}, 155456.
\bibitem{tbc1}Dolcetto, G.; Traverso Ziani, N.; Biggio, M.; Cavaliere, F.; Sassetti, M. Spin textures of strongly correlated spin Hall quantum dots. {\em Phys. Stat. Sol. (RRL)} {\bf 2013}, {\em 7}, 1059.
\bibitem{tbc2}Dolcetto, G.; Traverso Ziani, N.; Biggio, M.; Cavaliere, F.; Sassetti, M. Coulomb blockade microscopy of spin-density oscillations
and fractional charge in quantum spin Hall dots. {\em Phys. Rev. B} {\bf 2013}, {\em 87}, 235423.
\bibitem{alicea}Alicea, J. New directions in the pursuit of Majorana fermions in solid state systems. {\em J. Rep. Prog.} {\bf 2012}, {\em 75}, 076501.
\bibitem{giamarchi}Giamarchi T. Quantum Physics in One Dimension. {\em Oxford University Press} {\bf 2004}, {\em 9780198525004}.
\bibitem{franchini}Franchini, F. An Introduction to Integrable Techniques for
One-Dimensional Quantum Systems,{\em Lecture Notes in Physics}
Vol. 940 (Springer, Berlin {\bf 2017}).
\bibitem{xy}Porta, S.; Cavaliere, F.; Sassetti, M.; Ziani, N.T. Topological classification of dynamical quantum phase transitions in the xy chain. {\em Sci. Rep.} {\bf 2020}, {\em 10}, 642.
\end{thebibliography}
\end{document}